\documentclass[hyper]{JHEP3}
\usepackage{latexsym,amssymb,amsmath,epsfig,epic,eepic}
\usepackage{epstopdf}
\newcommand{\beq}{\begin{equation}}
\newcommand{\eeq}{\end{equation}}
\newcommand{\beqs}{\begin{eqnarray}}
\newcommand{\eeqs}{\end{eqnarray}} \newcommand{\tr}{\mathrm{tr\,}}

\newcommand{\nn}{\nonumber}

\newcommand{\ov}{\overline}

\title{Non-Perturbative Effects on a Fractional D3-Brane}

\author{\parbox{12cm}{ Gabriele Ferretti$^1$
and  Christoffer Petersson$^{1,2}$}
\\
\\
~\\
$^1$Department of Fundamental Physics \\
Chalmers University of Technology, 412 96 G\"oteborg, Sweden\\

\vspace{0.1cm}
$^2$PH-TH Division, CERN CH-1211 Geneva, Switzerland
\vspace{0.5cm}

\email{ferretti@chalmers.se, chrpet@chalmers.se}\\}

\abstract{In this note we study the $\mathcal{N}=1$ abelian gauge theory on the world volume of a single fractional D3-brane. In the limit where gravitational interactions are not completely decoupled we find that a superpotential and a fermionic bilinear condensate are generated by a D-brane instanton effect. A related situation arises for an isolated cycle invariant under an orientifold projection, even in the absence of any gauge theory brane. Moreover, in presence of supersymmetry breaking background fluxes, such instanton configurations induce new couplings in the 4-dimensional effective action, including non-perturbative contributions to the cosmological constant and non-supersymmetric mass terms.   }

\preprint{}
\begin{document}
\setcounter{section}{0}

\renewcommand{\thefootnote}{\arabic{footnote}}
\setcounter{footnote}{0} \setcounter{page}{1}


\section{Introduction}

The construction of the instanton action by means of string theory~\cite{Polchinski:1994fq,Witten:1995gx,Douglas:1995bn,Witten:1996bn,Green:2000ke,Billo:2002hm} has helped elucidating the physical meaning of the ADHM construction~\cite{Atiyah:1978ri} and allowed for an explicit treatment of a large class of non-perturbative phenomena in supersymmetric theories. Since string theory is a consistent enlargement of the field theory framework, we should not expect the effects of these instantons to be limited to those of their field theoretical counterpart. In fact, instantons in string theory, realized as wrapped Euclidean branes, give rise to additional effects, not only in the gravitational sector, but also in the gauge theories to which they couple.

In many cases of interest these effects arise by taking seriously the picture of the instanton as an independent wrapped brane and by allowing it to influence the dynamics of theories that would ordinarily not support a gauge instanton profile. Recently, instanton calculus in string theory has found many applications in attempts of constructing semi-realistic string vacua~\cite{Blumenhagen:2006xt,Ibanez:2006da,Cvetic:2007ku,Ibanez:2007rs,Blumenhagen:2007zk,Cvetic:2007qj,Ibanez:2008my,Kokorelis:2008ce} since Euclidean branes can give rise to couplings in the effective action that are forbidden to all orders in perturbation theory, such as Majorana masses for neutrinos and the $\mathbf{10}\times\mathbf{10}\times\mathbf{5}$ Yukawa coupling in GUT $SU(5)$.

In this note we analyze simple brane configurations which allow for various D-brane instanton effects.
In particular we focus on the $\mathcal{N}=1$ world volume theory of a single space-filling fractional D3-brane probing a singularity. Such a pure $U(1)$ gauge theory corresponds to the limiting case between gauge theories that admit ordinary gauge instanton effects and those that admit instanton effects which do not have an obvious interpretation in terms of ordinary field theory. By embedding this seemingly trivial theory within string theory, we are provided with a UV complete version which turns out to have several non-trivial features. First of all we note, from the point of view of both open and closed strings, that this pure $U(1)$ gauge theory seems to exhibit an asymptotically free  running of the gauge coupling constant at high energies. Then, by evaluating the moduli space integral, we find that a non-perturbative superpotential is generated by a D-instanton effect. Moreover, we set up and perform the calculation concerning the corresponding fermionic bilinear condensate and find it to be non-vanishing in a one-instanton background.

These results are of course in contrast with the vanishing results one obtains in standard commutative pure abelian gauge theory, and indeed we only find non-vanishing results in the limit where gravity is not completely decoupled and the field theory/ADHM interpretation is abandoned. The idea of working in the string theory limit in order to obtain non-vanishing instanton corrections to the 4-dimensional effective action was also used, for example, in \cite{Green:2000ke}. However, while that paper discussed D-instanton contributions to higher derivative terms in the $\mathcal{N}=4$ abelian gauge theory on a D3-brane in flat space, we will be concerned with fractional D-instanton contributions to the superpotential in the $\mathcal{N}=1$ abelian gauge theory on a fractional D3-brane at a singularity.

Analogous arguments can be applied to the case of an isolated vanishing 2-cycle which is invariant under an orientifold projection. Although no space-filling D-branes are wrapping the cycle and there is no notion of any gauge dynamics, we can still have a well-defined instanton action and moduli space integral. Also in this $Sp(0)$ gauge theory we find that a non-perturbative superpotential is generated by a wrapped Euclidean D1-brane (ED1-brane). Since this procedure in general induces an explicit dependence on several  of the resolved 2-cycle volumes in the superpotential, it is of interest in the context of moduli stabilization in type IIB flux compactifications \cite{Kachru:2003aw}.

The results we find agree with previous arguments that have been put forward in the context of geometric transitions and matrix models \cite{Aganagic:2003xq,Intriligator:2003xs}\footnote{See \cite{GarciaEtxebarria:2008iw} for a recent discussion concerning the relation between matrix models and D-brane instantons. The arguments in that paper that involve instanton generated superpotentials are related and in agreement with the corresponding results found in this note, but the derivations are different.}. These papers argued that a Veneziano-Yankielowicz superpotential should be present at low energies in UV complete versions of pure $U(1)$ and $Sp(0)$ theories. The reason is because residual instanton effects arise along the Higgs branch of these theories after brane-antibrane pairs have been added, and the gauge group has been embedded into a supergroup. It is interesting that these effects can also be explained by a direct  D-instanton computation. 

Finally, we discuss how these low rank gauge theories are affected when we turn on some background fluxes that induce soft supersymmetry breaking mass terms in the 4-dimensional theory. Since these fluxes also induce new couplings in the effective instanton action we obtain non-vanishing instanton corrections to the 4-dimensional effective action from configurations that give a vanishing contribution in absence of fluxes. We comment on configurations that give a non-perturbative contribution to the cosmological constant and also on configurations where both supersymmetric and non-supersymmetric mass terms are generated by a D-instanton effect. \\

\section{D-Instanton Effects in \boldmath{$\mathcal{N}=1$} World Volume Theories}

Consider a generic, local, $\mathcal{N} = 1$ IIB brane configuration on $\mathbb{R}^{3,1} \times K_6$. By ``local" we mean that we are considering only a small region of $K_6$ where the branes are present, ignoring global issues. Depending on the position of the D-brane instanton, relative to the space-filling branes, we distinguish between two non-trivial possibilities:\\

\noindent $\bf{Case~A:}$ The instantonic D-brane wraps a cycle upon which more than one space-filling D-brane are also wrapped. In this case the ED-brane can be interpreted as an ordinary gauge instanton. If the matter content allows it, such configurations can generate Affleck-Dine-Seiberg (ADS)~\cite{Affleck:1983mk,Taylor:1982bp,Akerblom:2006hx} superpotentials which schematically have the structure:
\begin{equation}
\label{bADSNc}
W^{\mathrm{np}}=  \frac{\Lambda^b}{\Phi^{b-3}},  \quad b>0,
\end{equation}
where $\Lambda$ is the dynamically generated scale and its exponent corresponds to the coefficient of the one-loop $\beta$-function. The  expression $\Phi^{b-3}$ denotes a generic gauge invariant combination of the chiral matter fields charged under the gauge group where the instanton resides. \\

\noindent $\bf{ Case ~B:}$ The ED-brane wraps a cycle which is either occupied by a single space-filling D-brane, or is unoccupied but invariant under an orientifold projection. In both cases the wrapped ED-brane can not be directly interpreted as an ordinary gauge instanton\footnote{See however \cite{Aharony:2007pr,Krefl:2008gs,Amariti:2008xu} for a discussion on how some of these instanton effects can be seen as a strong coupling effect by using Seiberg dualities.}.
However, such configurations may still give rise to a term in the superpotential if the matter content of the other nodes allows it 
\cite{Blumenhagen:2006xt,Ibanez:2006da,Florea:2006si,Bianchi:2007fx,Argurio:2007vqa,Bianchi:2007wy,Aganagic:2007py,GarciaEtxebarria:2007zv,Petersson:2007sc,Argurio:2008jm,Kachru:2008wt}. Such terms are polynomial in the matter fields since they arise only from the integration over the fermionic zero-modes and we write them schematically as
\begin{equation}
\label{bbADSNc}
W^{\mathrm{np}}=  \Lambda^{3-n} \Phi^{n},  \quad n \geq 0.
\end{equation}
Here, $3-n$ no longer corresponds to the coefficient of the beta function for any of the nodes to which $\Phi$ couples and the dimensionful constant $\Lambda$ is not the dynamical scale for these nodes. However, as will be discussed in the next section, $\Lambda^{3-n}$ is well-defined in terms of the D-brane instanton action and the dimension $3-n$ can be understood in terms of non-vanishing open string one-loop amplitudes. \\

\noindent We will also see in the next section that the $n=3-b=0$ case arises as an interesting limiting case of both these types of configurations. For this case, since there are no (non-vanishing vacuum expectation values of) chiral superfields connected to the instanton node, we need another way to introduce a finite scale in order to smoothen out the instanton moduli space singularities. This scale is naturally provided if we keep the string scale finite, and thus refrain from taking the strict field theory/ADHM limit. Even though we choose to work in a simple orbifold setting, the more general toric case can straightforwardly be inferred from this construction, using for example the prescriptions given in \cite{Argurio:2008jm,Kachru:2008wt}.

\subsection{Fractional D3-branes at an Orbifold Singularity}

We will now illustrate the physics outlined above by considering a
$\mathbb{C}^3 /\mathbb{Z}_2 \times \mathbb{Z}_2$ orbifold as an example \cite{Berkooz:1996dw}. Let us denote by $(N_1, N_2, N_3, N_4)$ a configuration with $N_i $ space-filling fractional branes D3${}_i$ at nodes $i=1,2,3,4$ in the quiver. On the world volume of these fractional D3-branes we obtain a (non-chiral) gauge theory with gauge group $\prod U(N_i)$ and with chiral superfields, $\Phi_{ij}$ with $i\neq j$, transforming in the bifundamental representations. Throughout this note we will only consider a single fractional D(-1)-instanton at node 1, denoted by D(-1$)_1$.

Let us first recall the zero mode structure of a D(-1$)_1$-instanton in such a system (see~\cite{Argurio:2007vqa} for details). In the neutral sector, consisting of modes of the open string beginning and ending on the D(-1$)_1$-instanton, there are 4 bosonic zero modes $x^\mu$ along with 3 auxiliary modes $D^c$ and 4 fermionic modes $\theta^\alpha$ and $\lambda^{\dot\alpha}$. In the charged sector, consisting of massless modes charged under the 4-dimensional gauge groups, there are 4$N_1$ bosonic moduli $\omega_{\dot\alpha}$, $\bar\omega_{\dot\alpha}$ from the strings stretching between the D(-1$)_1$-instanton and the $N_1$ D3${}_1$-branes. Furthermore, in the charged sector there are $2N_i$ fermionic modes $\mu_{i1}$, $\bar\mu_{1i}$ from the open strings stretching between the D(-1$)_1$-instanton and the $N_i$ D3${}_i$-branes at node $i$.

From the scaling dimension of the moduli fields (see e.g.~\cite{Billo:2002hm}) we obtain the dimension of the measure for the moduli space integral corresponding to this instanton configuration,
\begin{eqnarray}
\label{dimmeas2}
\Big[ d \{ x, \theta,  \lambda,D ,\omega , \ov\omega , \mu ,\ov\mu \Big] &=& M_s^{-(n_{x } -\frac{1}{2}n_{\theta} +\frac{3}{2}n_{\lambda} -2 n_{D}+ n_{\omega,\ov\omega} - \frac{1}{2}n_{\mu , \ov\mu } )} \nn \\
&=& M_{s}^{-( n_{\omega,\ov\omega} - \frac{1}{2}n_{\mu , \ov\mu } )} = M_{s}^{-(3N_1-N_2-N_3-N_4)}~.
\end{eqnarray}
In order to obtain a dimensionless term in the effective 4-dimensional action we need to have a prefactor for the moduli space integral that compensates for the dimension in (\ref{dimmeas2}). By also taking into account the contribution from the complexified vacuum instanton disk amplitude \cite{Polchinski:1994fq}, given by (minus) the node 1 instanton classical action $S^{ED1}_{1}=-2\pi i \tau_1$, we conclude that the prefactor should have the following structure
\begin{equation}
\label{pref}
\Lambda^{3N_1-N_2-N_3 -N_4}=M_{s}^{3N_1-N_2-N_3 -N_4}~e^{2\pi i \tau_1} \, .
\end{equation}
In this expression we refer to the general situation, away from the orbifold limit, where $\tau_1$ is the complexified volume of the node 1 resolved 2-cycle $\Sigma_1$ in $K_6$ upon which the ED1-brane is wrapped,
\begin{equation}
\label{tau1}
\tau_1 =
\frac{1}{4\pi^2 \alpha'}\int_{\Sigma_1}\Big[ C_2 +i e^{-\phi} \sqrt{\det g} \Big] ~.
\end{equation}
Here $C_2$ is the RR 2-form gauge potential, $\phi$ is the dilaton and $g$ is the string frame metric pulled back onto the world volume of the ED1-brane. Note that (\ref{pref}) and (\ref{tau1}) are well-defind even in the case when there are no spacefilling D5-branes wrapped on $\Sigma_1$, although in that case there is no 4-dimensional gauge coupling constant to which we can relate $\tau_1$. 

\vspace{.3cm}
\noindent \emph{One-loop corrections}
\vspace{.2cm}

\noindent The dimension of the prefactor of the D(-1$)_1$-instanton amplitude (\ref{pref}) can also be obtained by studying one-loop fluctuations around the instanton \cite{Akerblom:2006hx,Billo:2007sw,Billo:2007py}. A non-vanishing dimension corresponds to non-vanishing annulus\footnote{In the presence of orientifolds one must also take into account M$\ddot{\mathrm{o}}$bius vacuum amplitudes between the D(-1$)_1$-instanton and the orientifold.} vacuum amplitudes with one end on the D(-1$)_1$-instanton and the other end on one of the D$3_i$-branes. The massless modes circling the loop give rise to a logarithmic correction to the tree level vacuum D(-1)${}_1$ disk amplitude, and the sum of the coefficients of these corrections is precisely given by $3 N_1 - N_2 - N_3 - N_4$, in agreement with (\ref{pref}).

In the case when there are space-filling D5-branes wrapped on the same 2-cycle as the instanton we can relate this effect to the gauge coupling constant, $\tau_1 = \frac{\theta_1}{2\pi}+i \frac{4\pi}{g^{2}_{1}}$, for the world volume gauge theory of the D5-branes at node 1. This implies that the logarithmic corrections to the instanton action can now be identified with logarithmic corrections to the gauge coupling constant, and hence the dimension of the D-instanton amplitude prefactor can be identified with the one-loop $\beta$-function coefficient $b_1$ for the coupling constant $g_1$ of the gauge group at node 1. Furthermore, the dimension of the prefactor can now alternatively be obtained by considering one-loop amplitudes between spacefilling D5-branes with two gauge field vertex operators inserted along the boundary of the D5-branes at node 1 \cite{Akerblom:2006hx,Billo:2007sw,Billo:2007py}.

The one-loop $\beta$-function coefficient can also be obtained on the closed string side from the dual supergravity solution for fractional D$3$-branes at a $\mathbb{C}^3 /\mathbb{Z}_2 \times \mathbb{Z}_2$ orbifold singularity, given in~\cite{Bertolini:2001gg,Imeroni:2003cw}. By expanding the square root in the Dirac-Born-Infeld action to quadratic order in the gauge field we get the prefactor of the gauge kinetic term, and thereby an expression for the gauge coupling constant $g_1$ of the D$3_1$-brane world volume theory. This expression for $g_1$ incorporates the twisted scalars that correspond to the flux of the NS-NS $B_2$-field through the vanishing 2-cycles of the orbifold geometry. Since each type of fractional D$3$-brane is charged under all the three twisted sectors, they act as sources for all the twisted scalars and induce a logarithmic profile for them. By inserting this supergravity solution into the prefactor of the gauge kinetic term we recover the same logarithmic behavior as above\footnote{The reason why the result we obtained from circling the massless open string modes in the annulus calculation can be precisely mapped to the tree level result for the massless closed (supergravity) modes is because of the absence of threshold corrections from the massive modes to the prefactor of the gauge kinetic term~\cite{DiVecchia:2005vm}. Therefore, the infrared logarithmic divergence in the closed string tree-level channel due to the twisted tadpoles is exactly reflected in the open string one-loop channel as an ultraviolet logarithmic divergence due to the lack of conformal invariance (unless $N_1=N_2=N_3=N_4$) in the world volume theory.}.

The key point for our purposes is that all of these procedures give us an expression for the one-loop $\beta$-function coefficient for the gauge coupling constant $g_1$ of the D$3_1$-branes that is valid for $N_1=1$ as well as $N_1>1$. For the case we will be mostly interested in later on, when $N_1=1$ and $N_2=N_3=N_4=0$, we get that $b_1=3$, indicating that the abelian world volume theory on the single D$3_1$-brane exhibits an asymptotically free behavior at high energy. This non-vanishing coefficient of course does not agree with the vanishing result one obtains from an ordinary $\mathcal{N}=1$ pure abelian gauge theory\footnote{Note that the result $b_1=3$ does however agree with the result one finds for a pure $\mathcal{N}=1$ abelian gauge theory defined on a noncommutative background \cite{Martin:1999aq,Khoze:2000sy}.} and we interpret it as being due to the stringy UV completion.

\vspace{.3cm}
\noindent \emph{Instanton generated superpotentials}
\vspace{.2cm}

\noindent The cases $A$ and $B$ discussed around (\ref{bADSNc}) and (\ref{bbADSNc}) are known to arise in this particular orbifold and can both be induced by a D(-1$)_1$-instanton for the following surrounding of space-filling fractional D3-branes:\\

\noindent $\bf{ Case~ A:}$ $(N_1,N_2,N_3,N_4) = (N+1, N, 0, 0)$. The gauge group is $U(N+1)_1\times U(N)_2$ and we have chiral fields $\Phi_{12}$ and $\Phi_{21}$ transforming in the bifundamental representation. In this configuration a non-perturbative superpotential is generated by a D(-1$)_1$-instanton effect:
\begin{equation}
\label{ADSNc}
W^{\mathrm{np}}_{A}=  \frac{\Lambda^{3+2N }_{A}}{\det [\Phi_{21} \Phi_{12}]},
\end{equation}
where $\Lambda_{A}$ is the dynamical scale of $U(N+1)_1$ and the one-loop $\beta$-function coefficient is correctly given by $b_{1}^{A}=3N_1 - N_2=3+2N$.\\

\noindent $\bf{ Case~ B:}$ $(N_1,N_2,N_3,N_4) = (1, N, N, 0)$. The gauge group is $U(1)_1\times U(N)_2\times U(N)_3$ and the following non-perturbative superpotential is generated by the D(-1$)_1$-instanton~\cite{Petersson:2007sc}:
\begin{equation}
\label{hej}
W^{\mathrm{np}}_{B}=  \Lambda^{3-2N}_{B} \det [\Phi_{32} \Phi_{23}].
\end{equation}
Here $\Phi_{23}$ and $\Phi_{32}$ are in the bifundamental of the two $U(N)$ factors but
$\Lambda_B$ does not correspond to the dynamical scale of either of them. Instead, as discussed above, $\Lambda_B$ denote the fact that (\ref{hej}) is a D(-1$)_1$-instanton amplitude and $b_{1}^{B}=3 - N_2-N_3=3-2N$.

\vspace{.3cm}
\noindent \emph{Instanton-generated condensates}
\vspace{.2cm}

\noindent From the general relation between gaugino condensates and low energy effective superpotentials,
\begin{equation}
\label{rel}
\left<\tr [\Lambda^{\alpha}\Lambda_{\alpha}]  \right> \approx \frac{1}{b_1}\Lambda \frac{\partial}{\partial \Lambda} \left< W^{\mathrm{np}} \right>,
\end{equation} 
where $\Lambda^{\alpha}$ is the gaugino of the vector multiplet at node 1, we expect that the superpotentials generated in both case $A$ and $B$ are in one-to-one correspondence with the formation of a vacuum expectation value in a D(-1$)_1$-instanton background. Let us perform a simple counting of fermionic zero modes in order to see which type of condensates we should expect. Moreover, let us discuss the case when we have placed an arbitrary number of $k_1$ D(-1$)_1$-instantons at node 1.  

We first of all note that there is an equal number of massless fermionic modes, from the various types of open strings with at least one end attached to one of the $k_1$ D(-1$)_1$-instantons, for both case $A$ and $B$ since there is an equal number of fractional D3-branes in both configurations, although they are of type $(N+1,N, 0,0)$ in case $A$ and of type ($1,N,N,0$) in case $B$. This gives us the dimension of the fermionic part of the instanton moduli space \cite{Bianchi:2007wy} for both case $A$ and $B$, \footnote{Remember that we subtract the number of $\lambda$'s since they act as Lagrange multipliers enforcing the fermionic ADHM-constraints.}
\begin{equation}
\label{fermcount}
\mathrm{dim}\left[ \mathcal{M}_F \right]= n_{\theta} +n_{\mu , \ov\mu } -n_{\lambda}= n_{\mu , \ov\mu }=2k_1+4Nk_1 ~,
\end{equation}  
which we will now compare to the number of fermionic zero modes required by the following two types of condensates:\\

\noindent $\bf{ Case~ A:}$ From the relation (\ref{rel}) for case $A$ we get that the following condensate is formed in a D(-1$)_1$-instanton background \cite{Affleck:1983mk,Taylor:1982bp}, 
\begin{equation}
\label{ADScon}
\left<\tr [\Lambda^{\alpha}\Lambda_{\alpha}] \det\left[ \Phi_{21}\Phi_{12}\right] \right> = \Lambda^{3+2N}_{A},
\end{equation} 
where we have multiplied both sides of (\ref{rel}) with $\det\left[ \Phi_{21}\Phi_{12}\right]$. Since each gaugino soaks up one fermionic zero mode and (the scalar component of) each chiral superfield soaks up two, we need an instanton background with $2+4N$ fermionic zero modes, which agrees with (\ref{fermcount}) for $k_1=1$. If we instead were to place $k_1>1$ D(-1$)_1$-instantons at node 1, the condensate (\ref{ADScon}) would require the presence of  $\mathrm{dim}_A \left[ \mathcal{M}_F \right] =2+2(k_1 b_{1}^{A} -3)$ fermionic zero modes, which does not agree with (\ref{fermcount}). Thus, for $k_1>1$ we do not expect a condensate of type (\ref{ADScon}) to be generated.\\

\noindent $\bf{ Case~ B:}$ For case $B$, the relation (\ref{rel}) indicates that the following condensate should be formed in a D(-1$)_1$-instanton background, 
\begin{equation}
\label{U1con}
\left<\Lambda^{\alpha}\Lambda_{\alpha} \right> = \Lambda^{3-2N}_{B}~\left<\det\left[ \Phi_{32}\Phi_{23}\right]\right>,
\end{equation}
where we have removed the trace since $\Lambda^{\alpha}\Lambda_{\alpha}$ in (\ref{U1con}) refers to the abelian fermions in the $U(1)$ vector multiplet at node 1. If we were to consider the more general case with $k_1$ D(-1$)_1$-instantons, the condensate (\ref{U1con}) would require the presence of $\mathrm{dim}_B \left[ \mathcal{M}_F \right] =2+2(3-k_1 b_{1}^{B} )$ fermionic zero modes. Hence, since this dimension only agrees with (\ref{fermcount}) for $k_1=1$ it is only in a one-instanton background we expect a condensate of type (\ref{U1con}) to be generated.\\

\noindent In section \ref{comp} we will show that these expectations are fulfilled by doing explicit D-instanton computations.

\subsection{Non-Perturbative Effects in Pure $U(1)$ Gauge Theory}

The limiting situation for both case $A$ and $B$, when $N=0$, corresponds to a pure $U(1)$ gauge theory. For our specific orbifold it is possible to interpolate between these two configuration by moving various ``$\mathcal{N}=2$ branes"~\cite{Franco:2005zu} in and out from infinity. For instance, starting from the configuration $A$ it is possible to move a fractional D3-brane of type (1,1,0,0) away from the singularity, implying a Higgsing of the theory to $U(N)_1 \times U(N-1)_2$. Further Higgsing leaves us with a pure $U(1)$ theory on the first node. Similarly, we can get to the $U(1)$ theory by successively removing fractional branes of type $(0,1,1,0)$ from configuration $B$. By using renormalization group matching and the fact that $b_1=3$ for the pure $U(1)$ case, we are led to believe that also in the limiting $N=0$ case, a non-perturbative superpotential with the following structure is generated,
\beq
    \label{ADS1}
    W^{\mathrm{np}}=  \Lambda^3.
\eeq
In a more general situation, quite independent from the orbifold we used, such a theory corresponds to the intermediate case between (\ref{bADSNc}) and (\ref{bbADSNc}) where $n=b-3=0$, on a cycle without chiral matter. Since (\ref{ADS1}) has the structure of a one-instanton amplitude it should be generated by a one-instanton effect on an isolated node, and if so, we must be able to calculate it using the D-instanton techniques. Although there are no instanton effects in standard pure abelian gauge theory\footnote{This is in contrast to a noncommutative pure abelian gauge theory which does have non-singular instanton solutions \cite{Nekrasov:1998ss}.}, in the next section we show that in this string theory realization we do generate (\ref{ADS1}) because of the incomplete decoupling of gravity. Moreover, in accordance with the discussion in the previous section and the relation (\ref{rel}), we also expect for the particular case when $N=0$ that the superpotential (\ref{ADS1}) is generated whenever the following condensate is formed in a D(-1$)_1$-instanton background,
\begin{equation}
\label{gaugecond}
\left<\Lambda^{\alpha}\Lambda_{\alpha} \right> = \Lambda^{3}~.
\end{equation}

Note that (\ref{ADS1}) and (\ref{gaugecond}) have the same structure as in the case of the usual non-abelian gaugino condensation for pure $\mathcal{N}=1$ SYM. In this case however, when the number of colors is greater than one, neither  (\ref{ADS1}) nor (\ref{gaugecond}) can be generated directly by a one-instanton effect since the one-loop $\beta$-function coefficient does not agree with the dimension of the superpotential or the condensate.

Also note that both (\ref{ADS1}) and (\ref{gaugecond}) are expected from the point of view of geometric transitions \cite{Gopakumar:1998ki,Cachazo:2001jy} and the Dijkgraaf-Vafa theory~\cite{Dijkgraaf:2002dh, Kraus:2003jf,Cachazo:2003kx}, (see also~\cite{Aganagic:2003xq,Intriligator:2003xs,Aganagic:2007py,GarciaEtxebarria:2008iw}). One might for example argue that even a single D5-brane, wrapped on one of the 2-cycles in the resolved geometry, should trigger a geometric transition\footnote{Note that even the fractional D3-branes at the $\mathbb{C}^3 /\mathbb{Z}_2 \times \mathbb{Z}_2$ orbifold singularity we are using here  are expected to deform the geometry in the IR~\cite{Imeroni:2003cw}.} resulting in a finite sized 3-cycle with a single unit of 3-form flux through it. This should then be reflected in the low energy effective theory by the presence of a  Veneziano-Yankielowicz (VY) superpotential~\cite{Veneziano:1982ah} for the glueball field $S \approx \tr \left[ \Lambda_\alpha\Lambda^\alpha \right]$, corresponding to the size of the 3-cycle:
\beq
    W_\mathrm{VY} = h(G) S \left( 1 - \log\frac{S}{\Lambda^3} \right),
\eeq
where $h(G)$ is the dual Coxeter number of the gauge group $G$. In a UV complete framework it is expected to have the following generalized definitions \cite{Aganagic:2003xq,Intriligator:2003xs},
\beq
     h(U(N)) = N, \quad h(Sp(N)) = N+1, \quad h(SO(N)) = N-2,
\eeq
valid for \emph{all} $N\geq 0$ and not just for those values corresponding to non-abelian gauge groups. For instance, one gets $h(U(1)) = 1 \not= h(SO(2))=0$ and $h(Sp(0)) = 1$.  Thus, this prescription tells us that a VY superpotential should be added in the $U(1)$ case, and hence that the fermion bilinear should acquire a non-vanishing vacuum expectation value, corresponding to (\ref{gaugecond}), in agreement with the fact that the 3-cycle in the IR regime acquires  a finite size. Moreover, when inserting this result back into the VY superpotential one obtains (\ref{ADS1}). In the following section we will show that these results for the $U(1)$ case (and the $Sp(0)$ case) can be precisely explained by a D-instanton effect on a 2-cycle upon which a single D5-brane (or an orientifold plane for the $Sp(0)$ case) is also wrapped.

Finally, one might be worried that the $U(1)$ vector multiplet is rendered massive at the string scale by its coupling to the background RR-fields. However, since this $U(1)$ is non-anomalous it is massless in the non-compact limit and only gets a mass upon compactification~\cite{Antoniadis:2002cs,Buican:2006sn,Conlon:2008wa}, which can be much smaller than the string scale. Our statements are applicable within this range of masses.

\subsection{Computation of the Superpotential and the Condensates}
\label{comp}

Having argued from many different points of view that contributions like (\ref{ADS1}) and (\ref{gaugecond}) are expected when the $U(1)$ node is embedded into a constistent stringy UV completion, we now proceed to an explicit computation using the corresponding instanton action.

The superpotential can be computed by evaluating the moduli space integral for the configuration with a D(-1)${}_1$-instanton and a D3${}_1$-brane,
\begin{equation}
\label{dimmeas}
 S^{\mathrm{4-d}}_{\mathrm{np}}= \int d^4 x d^2 \theta ~W_{\mathrm{np}}= \int d^4 x d^2 \theta\Big[\Lambda^{3} \int  d^2 \lambda^{\dot\alpha} d^3D^c d^2\omega_{\dot\alpha} d^2 \ov\omega^{\dot\alpha} d\mu_{11} d\ov\mu_{11} ~e^{-S^{\mathrm{0-d}}_{\mathrm{moduli}}}\Big],
\end{equation}
where the instanton action for the moduli fields is given by \cite{Green:2000ke,Billo:2002hm}, \footnote{As argued before from the fermionic zero mode counting, there is no contribution from configurations with more that one D(-1)${}_1$-instanton at node 1. This can be explicitly seen here since for $k_1 >1$ we would have $2k_{1}^{2}$ fermionic Lagrange multipliers $\lambda^{\dot\alpha}$ but only $2k_1$ charged fermions $\mu_{11}$ and $\bar\mu_{11} $. Thus, when we integrate over the $\lambda^{\dot\alpha}$-modes we always get that each charged fermionic zero mode appear more than once in the measure and hence anti-commutes to zero.}
\begin{equation}
\label{mod}
S_{\mathrm{moduli}}^{\mathrm{0-d}}=
 \frac{1}{2g_{0}^{2}}  (D^c)^2+i D^c\!\left( \bar\omega^{\dot \alpha}
(\tau^c)^{\dot\beta}_{\dot\alpha} \omega_{\dot\beta} \right)+i \left(\bar\mu_{11} \omega_{\dot\alpha} +
\bar\omega_{\dot\alpha} \mu_{11} \right)\! \lambda^{\dot\alpha},
\end{equation}
and the dimensionful 0-dimensional coupling constant reads $1/g_{0}^{2} =4\pi^3 {\alpha'}^2 /g_s$. Note that the prefactor $\Lambda^{3}$ in (\ref{dimmeas}) already saturates the dimension of a superpotential term. Thus, from dimensional analysis we can conclude that the result of the integral must have the structure of (\ref{ADS1}), up to a dimensionless constant.  Let us show that this dimensionless constant is non-zero.

The $\lambda^{\dot\alpha}$-variables only appears linearly and when we integrate them out we bring down two fermionic $\delta$-functions in the measure, enforcing the fermionic ADHM-constraints. From the product of these two $\delta$-functions we get a cross-term that contain both the $\mu_{11}$ and the $\ov\mu_{11}$ variable which we integrate out. We are then left with the following bosonic integral:
\begin{eqnarray}
\label{mum}
W_{\mathrm{np}}&=&\Lambda^{3} \int  d^3D^c d^2\omega_{\dot\alpha} d^2 \ov\omega^{\dot\alpha}~ \big(  \ov\omega^{\dot\alpha}\omega_{\dot\alpha}\big) ~e^{ -\frac{1}{2g_{0}^{2}}  (D^c)^2-i D^c\!\left( \bar\omega^{\dot \alpha}
(\tau^c)^{\dot\beta}_{\dot\alpha} \omega_{\dot\beta} \right)} \nn \\
&=&\Lambda^{3} \int  d^3D^c d^4 y
~\big( \vec{y}\cdot \vec{y}  \big)~e^{-2i (y_1 y_3+y_2 y_4)D^1- \frac{1}{2g_{0}^{2}} (D^1)^2 }
~
 \nn \\
&&\qquad \quad \times ~ e^{-2i (y_1 y_4-y_2y_3)D^2- \frac{1}{2g_{0}^{2}} (D^2)^2} ~~ e^{-i (y_{1}^{2}+y_{2}^{2}-y_{3}^{2}-y_{4}^{2})D^3- \frac{1}{2g_{0}^{2}} (D^3)^2 }~,
\end{eqnarray}
where $\omega_{\dot{1}} = y_1+i y_2$ and $\omega_{\dot{2}}=y_3+iy_4$. If we were to take the field theory/ADHM limit, $g_0 \to \infty$, or equivalently, $\alpha'\to 0$ with $g_s$ fixed, then the quadratic $(D^c)^2$-terms would vanish and the  $D^c$-fields would act as Lagrange multipliers, enforcing the ordinary ADHM constraints and hence set the instanton size, $\rho^2 = \ov\omega^{\dot\alpha}\omega_{\dot\alpha}=\vec{y}\cdot \vec{y}$, to zero. In this case we would not get any contribution to the superpotential. \footnote{One way to prevent the instanton from shrinking to zero size and smoothen out the moduli space singularity is to add a Fayet-Iliopoulos term $iD^c \xi^c$ to the effective instanton action (\ref{mod}). This term is added when the gauge theory is defined on a non-commutative background and it implements a deformation of the bosonic ADHM constraints~\cite{Nekrasov:1998ss}.}

Thus, we refrain from taking the limit $\alpha'\to 0$ and thereby give up the ordinary ADHM instanton moduli space interpretation, implying that from here on we are considering a true D(-1)${}_1$-instanton effect in the D3${}_1$-brane world volume theory. We can now use the fact that $\int e^{2bx-ax^2}dx=\sqrt{\pi/a}\,\,e^{b^2/a} $ for $a>0$ and obtain the following simple expression,
\begin{eqnarray}
\label{modint1}
W_{\mathrm{np}}&=&\Lambda^{3} \big(2\pi g_{0}^{2}\big)^{\frac{3}{2}} \int   d^4 y ~\Big( \vec{y}\cdot \vec{y}  \Big)~  ~e^{-\frac{g_{0}^{2}}{2}(\vec{y}\cdot \vec{y})^2}~\nn \\
&=&\Lambda^{3} \big(2\pi g_{0}^{2}\big)^{\frac{3}{2}}(\mathrm{vol}S^3) \int d\rho ~\rho^5 ~e^{-\frac{g_{0}^{2}}{2}\rho^4} \nn \\
&=&\Lambda^{3} 2 \pi^{4}
\end{eqnarray}
The trivial numerical constant can be absorbed into the definition of $\Lambda$. What is important is that the result is independent of $g_0$. This implies that the procedure of not decoupling gravity completely can be seen as a regularization which introduces a minimal scale, smoothens out the moduli space singularity and gives rise to a non-vanishing contribution to the superpotential.

\vspace{.3cm}
\noindent \emph{Computing the condensates}
\vspace{.2cm}

\noindent We can perform a similar computation to show the formation of a corresponding condensate involving the fermionic bilinears. We will do this in all generality, recovering (\ref{ADScon}) and (\ref{U1con}) for case $A$ and $B$ and also (\ref{gaugecond}) for the $U(1)$ case.

For this we need the instanton profile for the 4-dimensional gauginos. The profile for any of the  4-dimensional fields can be obtained by computing tree level amplitudes on mixed disks with one vertex operator insertion for a gauge theory field and the remaining insertions for moduli fields \cite{Green:2000ke,Billo:2002hm}. Although such a mixed disk amplitude has multiple insertions from the point of view of the worldsheet it should be thought of as a 1-point function from the point of view of the 4-dimensional gauge theory. The non-dynamical moduli fields merely describe the non-trivial instanton background on which the dynamical 4-dimensional fields depend. The instanton profile is then obtained by multiplying the mixed disk amplitude with a massless propagator and taking the Fourier transform~\cite{Billo:2002hm}.

We begin by considering the case where we have placed $N_1$ D$3_1$-branes at node 1, together with the D(-1$)_1$-instanton. The gaugino has tadpoles on mixed disks with either $\omega_{\dot\alpha}$ and $\bar\mu$ moduli insertions or with $\bar\omega_{\dot\alpha}$ and $\mu$ insertions. In addition to the profile contribution these amplitudes give rise to we also get a contribution from when we act with the supersymmetry generators that were broken by the D(-1$)_1$-instanton \cite{Green:2000ke}. This shifts the zero modes that correspond to the broken supersymmetries and thereby introduces an extra term in the gaugino profile that depends explicitly on $\theta^{\alpha}$. From this analysis we obtain an expression for a pair of gauginos with the following structure \cite{Dorey:2002ik},
\begin{equation}
\label{ll}
\tr [\Lambda^{\alpha}\Lambda_{\alpha}] = \frac{\rho^4\theta^{\alpha}\theta_{\alpha}}{\big[ (X-x)^2 +\rho^2 \big]^4}+\cdots ~,
\end{equation}
where $X^\mu$ is the space-time coordinate while $x^\mu$ still denotes the position and $\rho$ the size of the instanton. The ellipses denote terms with less powers of $\theta^\alpha$ that will not be important for our purposes. 

The expression (\ref{ll}) for the pair of gauginos in terms of the unconstrained moduli fields can now be inserted into the moduli space integral yielding
\begin{equation}
\label{ll1}
\left<\tr [\Lambda^{\alpha}\Lambda_{\alpha}]\right>=\Lambda^{b_1}  \int  d \{ x , \theta ,\lambda,D,\omega,\ov\omega,\mu,\ov\mu \} ~\tr [\Lambda^{\alpha}\Lambda_{\alpha}] ~e^{-S^{\mathrm{0-d}}_{\mathrm{moduli}}}~.
\end{equation}
As usual, $x^{\mu}$ and $\theta^{\alpha}$ correspond to the supertranslations broken by the D(-1$)_1$-instanton and do not appear explicitly in the instanton action $S^{\mathrm{0-d}}_{\mathrm{moduli}}$. They do however appear in the expression for the gaugino pair and we can use (\ref{ll}) when performing the integrals over these two variables,
\begin{equation}
\label{xtint}
\int d^4 x d^2 \theta~ \tr [\Lambda^{\alpha}\Lambda_{\alpha}]= \int d^4 x \frac{\rho^4}{\big[ (X-x)^2 +\rho^2 \big]^4}=\frac{\pi^2}{6},
\end{equation}
where we see that the factors of $\rho$ cancel off and we simply get a dimensionless constant which we can absorb in the prefactor $\Lambda^{b_1}$ of the remaining integral
\begin{equation}
\label{ll2}
\left< \tr [\Lambda^{\alpha}\Lambda_{\alpha}] \right>=\Lambda^{b_1}  \int  d \{ \lambda,D,\omega,\ov\omega,\mu,\ov\mu \} ~e^{-S^{\mathrm{0-d}}_{\mathrm{moduli}}}~.
\end{equation}
Now, the crucial point is that the integral that remains to be calculated in (\ref{ll2}) is precisely the integral one evaluates when computing the superpotential correction generated by the instanton configuration.

For case $A$ and the condensate (\ref{ADScon}), corresponding to the ADS superpotential (\ref{ADSNc}), the zero mode structure and the effective instanton action for a configuration with a \hbox{D(-1$)_1$}-instanton and a fractional D3-brane with rank assignment $(N+1,N,0,0)$ is given in \cite{Argurio:2007vqa}. The result  of the moduli space integral is given in (\ref{ADSNc}) and when we multiply both sides of (\ref{ll2}) with the product of chiral superfields we recover (\ref{ADScon}).

Similarly, for case $B$ and the condensate (\ref{U1con}), corresponding to the superpotential (\ref{hej}), the moduli space integral for a configuration with a D(-1$)_1$-instanton and a fractional D3-brane with rank assignment $(1,N,N,0)$ is given in \cite{Petersson:2007sc}.

Finally, in the limiting case $N=0$ ($b_1 = 3$) for the pure $U(1)$ theory the relevant integrals were performed above, starting from (\ref{dimmeas}), and the result we found implies that, for $\alpha' \neq 0$, (\ref{gaugecond}) is generated by a D(-1$)_1$-instanton effect.

\subsection{The Pure $Sp(0)$ Case}

It is straightforward to generalize the above considerations to the case when orientifolds are present. The specific example of $\mathbb{C}^3 /\mathbb{Z}_2 \times \mathbb{Z}_2$ was treated in detail in~\cite{Argurio:2007vqa}. For one particular choice of O3-plane, all the gauge groups turn into groups of symplectic type and the fields $\Phi_{ij}$ and $\Phi_{ji}$ get identified. Moreover, the two conjugate sectors among the charged zero modes also get identified while $\lambda^{\dot\alpha}$ and the $D^c$ modes of the neutral sector are projected out in the one-instanton case by the O3-plane.

By again placing a D(-1$)_1$-instanton at node 1 we get by dimensional analysis of the moduli space measure that the dimension of the instanton prefactor $\Lambda^{\tilde{b}_1}$ should be $\tilde{b}_1=(n_{x }+n_{\omega})-(1/2)(n_{\theta} + n_{\mu  } )=(6+3N_1 -N_2-N_3-N_4 )/2$. In the case when there are no fractional D3-branes ($N_1 =N_2=N_3=N_4=0$) and hence no gauge dynamics we still have a well-defined D(-1$)_1$-instanton action and $\tilde{b}_1=3$. This non-vanishing dimension, due to the neutral zero mode structure, can be identified with the coefficient of a logarithmic correction from a non-vanishing M$\ddot{\mathrm{o}}$bius vacuum diagram with one end on the D(-1$)_1$-instanton and the other on the O3-plane \cite{Bianchi:2007wy}.

Case $\tilde{A}$ now requires a $(N,N,0,0)$ configuration since we expect an ADS superpotential for an $Sp(N)$ theory when there are $N$ flavors present~\cite{Intriligator:1995ne},
\begin{equation}
\label{ADSNcSp}
W^{\mathrm{np}}_{\tilde{A}}=  \frac{\Lambda^{3 + N}_{\tilde{A}}}{\det \Phi_{12} }~.
\end{equation}
Similarly, case $\tilde{B}$ is given by the configuration $(0,N,N,0)$ for which \cite{Argurio:2007vqa},
\begin{equation}
\label{ADS0Sp}
W^{\mathrm{np}}_{\tilde{B}}=  \Lambda^{3 - N}_{\tilde{B}} \det \Phi_{23}~.
\end{equation}
In order to recover (\ref{ADS0Sp}) from (\ref{ADSNcSp}) we can start from a $(N,N,0,0)$ brane and move $N$ (1,1,0,0) branes away from the orbifold fixed point in a transverse complex direction and then move $N$ (0,1,1,0) branes into the fixed point. In order for the renormalization group matching to continuously take us between case $\tilde{A}$ and case $\tilde{B}$ we must have that a superpotential is generated also for the case when $N=0$. 

It is obvious that the corresponding moduli space integral is well-defined and non-vanishing for the pure $Sp(0)$ case since the charged sector is empty for $N=0$ and the $\lambda^{\dot\alpha}$ and $D^c$ fields have already been projected out by the orientifold. Hence, there are no ADHM constraints, no integrals to perform and we can immediately verify that $W = \Lambda^3$, again in agreement with the discussion in \cite{Aganagic:2003xq,Intriligator:2003xs} about the $Sp(0)$ case.
In a more general setup we expect contributions of this type to arise whenever a cycle obeys the above conditions.
This phenomena is of interest when studying moduli stabilization since in this way we induce an explicit K$\ddot{\mathrm{a}}$hler moduli dependence in the superpotential without the need for any space-filling D-branes.

\section{Instanton Effects in Flux Backgrounds}

 Compactifications in the presence of background fluxes are of great relevance to string phenomenology in the context of moduli stabilization. It is thus important to understand the interplay between fluxes and effective interactions in the D-brane world volume theories \cite{Tripathy:2005hv,Martucci:2005rb,Bergshoeff:2005yp,Blumenhagen:2007bn}, such as flux-induced supersymmetry breaking terms  \cite{Camara:2003ku,Grana:2003ek} and instanton zero mode lifting \cite{GarciaEtxebarria:2008pi,Billo':2008pg,Billo':2008sp,Uranga:2008nh}.     
   
We will in this section follow the world sheet approach of \cite{Billo':2008pg,Billo':2008sp} and use the results and notation from those papers. In the first example, we turn on $G_3$-flux of type (0,3) which gives a soft supersymmetry breaking mass to the gravitinos \cite{Camara:2003ku,Grana:2003ek,Billo':2008pg}. Furthermore, this type of flux induces a coupling in the instanton action to the neutral $\lambda^{\dot\alpha}$ moduli fields, implying that a single fractional D(-1)-instanton contributes to the superpotential even without any fractional D3-branes or orientifolds. Then, we discuss backgrounds where we have turned on flux of type (3,0), which generically induces soft supersymmetry breaking mass terms for the 4-dimensional gauginos \cite{Camara:2003ku,Grana:2003ek,Billo':2008pg}. The effect of turning on (3,0) flux can be seen as giving a vacuum expectation value to the auxiliary $\theta^2$-component of the ``spurion" $\tau_1$ chiral superfield from (\ref{tau1}). We will ignore any kind of backreaction of the background geometry due to the presence of  fluxes.

\vspace{.3cm}
\noindent \emph{Turning on (0,3)-flux}
\vspace{.2cm}

\noindent Let us begin by considering an instanton configuration with one D(-1$)_1$-instanton as usual, but with no fractional D3$_1$-branes or orientifold planes. In this case we expect no superpotential to be generated since the two Grassmann variables $\lambda^{\dot\alpha}$ do not appear in the instanton action.\footnote{Recall that in the absence of space-filling D-branes, orientifolds and fluxes a D-instanton breaks 4 of the 8 background supercharges and therefore has too many neutral fermionic zero modes.} However, if we turn on some supersymmetry breaking (0,3)-flux, a coupling to these variables appears \cite{Billo':2008sp} and the moduli space integral becomes,
\begin{equation}
\label{tjoff}
W_{(0,3)}^{\mathrm{np}} = e^{2\pi i \tau_1} \int  d^3D^c d^2 \lambda^{\dot\alpha} ~e^{ -\frac{2\pi^3 \alpha'{}^2 }{g_s}  (D^c)^2 +i G_{(0,3)}\frac{2\pi^3 \alpha'{}^2}{\sqrt{g_s}} \lambda^{\dot\alpha}\lambda_{\dot\alpha} } \approx e^{2\pi i \tau_1} ~\frac{g_s G_{(0,3)}}{\alpha'}~ .
\end{equation}
From (\ref{dimmeas2}) we see that this moduli space measure is dimensionless (if one includes $x^\mu$ and $\theta^\alpha$ in the counting), implying that the prefactor should also be dimensionless\footnote{This agrees with the fact that the charged sector is empty,  there are no annulus diagrams  and hence no logarithmic corrections to the vacuum D(-1$)_1$ disk amplitude.} and be given only by $e^{2\pi i \tau_1}$. 

\vspace{.3cm}
\noindent \emph{Turning on (3,0)-flux}
\vspace{.2cm}

\noindent Let us now consider the configuration with a single D3${}_1$-brane and a D(-1)${}_1$-instanton, but in a background with (3,0) $G_3$-flux. The way we implement this background flux is by adding the following interactions to the effective instanton action in (\ref{mod}) \cite{Billo':2008pg,Billo':2008sp},
\begin{equation}
\label{flux}
S_{\mathrm{(3,0)}}^{0-\mathrm{d}}=2\pi i \bigg( \frac{ 2G_{(3,0)} }{ \sqrt{g_s} }\theta^{\alpha}\theta_{\alpha}\bigg) + i\sqrt{g_s}G_{(3,0)} \ov\mu_{11}\mu_{11}~.
\end{equation}
Note that since this particular type of flux does not induce any additional interactions for the $\lambda^{\dot\alpha}$ variables in (\ref{mod}) they still act as Lagrange multipliers and pull down the fermionic $\delta$-functions which soak up both $\mu_{11}$ and $\ov\mu_{11}$. Thus, the last term in (\ref{flux}) does not play any role in the integration of $\mu_{11}$ or $\ov\mu_{11}$.

However, by including the first term of (\ref{flux}) in the instanton action (\ref{mod}), we are given the opportunity to explicitly soak up the $\theta^{\alpha}$-variables as well. Hence, by using our derivation of (\ref{ADS1}) we are able to evaluate the moduli space integral in this flux background and obtain the following one-instanton generated term in the 4-dimensional effective action,
\begin{equation}
\label{fluxresult}
S_{\mathrm{np}}^{\mathrm{4-d}}\approx  \int d^4 x ~\Lambda^3 \frac{G_{(3,0)}}{\sqrt{g_s}}~.
\end{equation}
We can view the term in (\ref{fluxresult}) as a non-perturbative contribution to the cosmological constant in the effective theory in which supersymmetry is softly broken.

Let us finally consider a related configuration, $(1,1,1,0)$, where we have also placed fractional D3-branes at nodes 2 and 3. By turning on a background (3,0)-flux in this setting we are given two different opportunities to soak up the two $\theta^{\alpha}$ modes. If we do not make use of the flux induced terms in (\ref{flux}), then the D(-1)${}_1$-instanton gives rise to the following non-perturbative supersymmetric mass term \cite{Petersson:2007sc},
\beq
\label{mutem}
W^{\mathrm{np}}= \Lambda \Phi_{23} \Phi_{32}~.
\eeq
On the other hand, if we do make use of the first term in (\ref{flux}), only the lowest components $\phi$ of the chiral superfields $\Phi$ in (\ref{mutem}) survive and we are left with the following non-supersymmetric mass term,
\begin{equation}
\label{bmu}
S_{\mathrm{np}}^{\mathrm{4-d}} \approx \int d^4 x ~  \Lambda \frac{G_{(3,0)}}{\sqrt{g_s}} \phi_{23} \phi_{32}~.
\end{equation}
Note that the terms (\ref{mutem}) and (\ref{bmu}) are reminescent of $\mu$ and $B_{(\mu)}$ terms and moreover, that the flux-induced mass for the gauginos is given by, $m_g \approx \sqrt{g_s} G_{(3,0)}$. This suggests that these three observables might also be related in more realistic configuations where both fluxes and D-instantons are taken into account.

\section*{Acknowledgements}

It is a pleasure to thank R.~Argurio, M.~Bertolini, and A.~Lerda for ongoing discussions and collaborations on related issues. C.P. would like to thank A.~Uranga for many helpful discussions and also J.~F.~Morales, D.~Persson and E.~Witten for valuable conversations. The research of G.F. is supported in part by the Swedish Research Council (Vetenskapsr{\aa}det) contracts 622-2003-1124 and 621-2006-3337. The research of C.P. is supported by an EU Marie Curie EST fellowship.


\end{document}